\documentclass[twocolumn,10pt]{article}

\usepackage{epsfig}
\usepackage{amssymb,amsmath,amsthm,graphicx}
\usepackage{color}
\usepackage[colorlinks=true,urlcolor=blue,linkcolor=blue,citecolor=blue]{hyperref}

\usepackage[T1]{fontenc}
\usepackage[utf8]{inputenc}
\usepackage{authblk}

\usepackage{hyperref}

\title{Tunable Focusing by a Flexible Metasurface}
\author{Yair Z\'arate$^{*}$, Ilya~V.~Shadrivov \& David~A.~Powell}
\affil{Nonlinear Physics Centre and Centre for Ultrahigh-bandwidth Devices for Optical Systems (CUDOS), Research School of Physics and Engineering, The Australian National University, Canberra, ACT 0200 Australia}

\begin{document}

\maketitle


Metasurfaces represent the most promising class of metamaterials for applications, whereby arbitrary wavefront and polarisation control can be achieved using just a single sub-wavelength layer \cite{Grbic1,Capasso2}. Examples of demonstrated functionality include reflection and refraction into anomalous directions \cite{Sun1}, generation of beams with orbital angular momentum  and focussing \cite{Capasso1}. Essential to this functionality is the ability to have full control over phase of the reflected or transmitted waves, so that the full $2\pi$ range of phase response can be accessed. In practice, many applications require a broader range of phase values, for example a lens can modulate phase in a very large range, however for a fixed operating frequency these values can be ``wrapped'' back into the fundamental range $[-\pi, \pi]$, introducing a spatial discontinuity in  parameters, and in our example the lens reduces to a Fresnel lens. This phase wrapping becomes problematic if the functionality (e.g. focal length or steering angle) or the operating frequency need to be modified. In such cases, the discontinuity needs to be shifted to a different location on the surface, which is very challenging to implement for regular optics and in a metamaterial based upon tunable electronic inclusions \cite{Liang1,Jangyeol1}. In this paper we resolve this problem by incorporating our metasurface within a flexible elastic structure.
\begin{figure}[hb!]
\centerline{\includegraphics[width=\columnwidth,height=8cm]{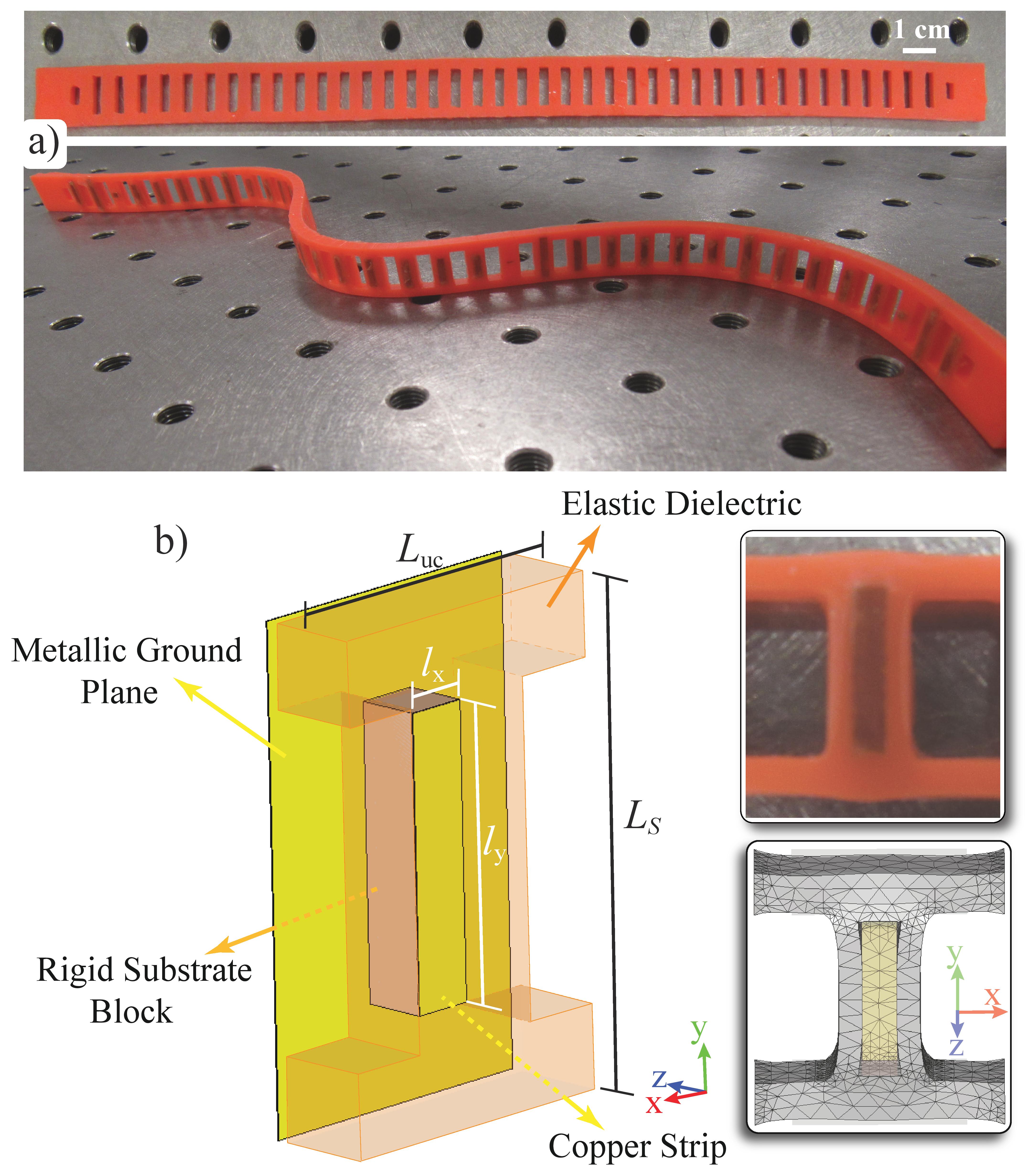}}
\caption{(a) Elastic metasurface designed to work in the microwave range. It is composed of an elastic dielectric holder in which electromagnetic resonators are inserted. (b) schematic of the building block, the inset shows the empirical test and the numerically calculated deformation of the unit cell after applying a $50\%$ elongation. }
\label{Fig-Setup}
\end{figure}

The incorporation of elastic materials with unconventional mechanical properties in the design of reconfigurable metamaterials has proven to be an appealing approach for the creation of tunable electromagnetic devices \cite{Smith1,Gupta1}. The great flexibility of elastic materials in terms of fabrication allows the creation of complex structures in which electromagnetic properties can be controlled by external stimulus\cite{Singer,Qiming1}. Such control becomes more effective when electromagnetic resonators are embedded in specially engineered elastic materials, where novel applications across the spectrum can be implemented by  harnessing the unique geometrical conformation that the surrounding elastic medium achieves under strain\cite{Walia1}. This integration enables the use of degrees of freedom in the design of optical devices based on metamaterials, allowing the creation of surfaces or three-dimensional materials with tunable electromagnetic responses that depend on the spatial configuration\cite{Nader1}. Such tunability is attained by exploiting the reconfigurable properties of the unit cells of electromagnetic-elastic metamaterials which can be accessed individually. Thereby, functionalities that rely on the collective behavior of the metamaterial components can be tuned externally. Applications for electromagnetic-elastic materials range from meta-molecules for tuning circular polarization of electromagnetic waves \cite{Yair}, bendable electronic circuits (stretchable electronics) \cite{Rogers2009}, which have proven to be a reliable platform for fabrication of electronic devices for being worn or implanted in living tissues \cite{Rogers2015_AdvFunMat, Rogers2015_NatBio}. Recently, a tunable metasurface on a stretchable substrate able to continuously manipulate the cross-polarized wavefront shape of circular polarized light has been proposed and experimentally demonstrated\cite{Agarwal}.

In the present work we report the experimental realization of tunable focusing by a flexible reflective metasurface with focal distance controlled by the applied strain. Our results are compared with theory and numerical simulations, showing good agreement. Although simple in design, our approach allows reliable control over the properties of the metasurface.

Metasurfaces are functional two-dimensional structured arrays of elements that can transform reflected and transmitted waves in an almost any arbitrary way. The elements of such metasurface are often made in the form of subwavelength resonators that control both amplitude and phase of electromagnetic fields. The concept of metasurfaces opened up new avenues for modifying amplitude, polarization and direction of light propagation.

{\em Design and Fabrication of the Flexible Metasurface.} Our focusing reflector is designed in the form of a flat metasurface composed of electric resonators embedded in a  patterned elastic holder (see Figure \ref{Fig-Setup}a), which is used as a platform to allow continuous elongation of the system\cite{Agarwal}. To fabricate the sample, we first produced the elastic holder with patterned slots for the electric resonators to be inserted. This has been cast by pouring a liquid A-silicone based rubber mixture, Heraeus Heraform Silicone Type A+B, into a teflon mold.  The electric resonators are made by cutting blocks of FR-4 printed circuit board. The rigid substrate shown in Figure \ref{Fig-Setup}b is the FR-4 with the copper strip. Then the electric resonators are placed in the silicone holder and glued using the same liquid silicone.

To determine the electromagnetic constant of the elastic dielectric we have produced a rubber disk of the same material of the elastic holder of the thickness of $10\,mm$  and with $29.5\,mm$ radius, and then placed it inside a waveguide of circular cross section of the same radius. Then we measured the scattering parameters using a vector network analyzer Rohde \& Schwarz ZVB $20$. Consequently, we adopted the approach  proposed in References \cite{Smith2002, Tretyakov2011}, based on the Nicolson-Ross-Weir method, to retrieve the electric permittivity as function of the transmission and the propagating mode inside the cylindrical waveguide. After evaluating the normalized experimental transmission coefficient in the relation for the electric permittivity we obtain that the electromagnetic parameter of the material are, $\epsilon_r=2.75$ and $\tan(\delta)=0.008$.

The flexible metasurface is backed by a fixed metallic ground plane. We fabricate the metasurface so that, when illuminated by a plane wave, the phase of the waves reflected by the surface is the same at the focal point, and almost full reflection of energy is achieved. We first investigate the electromagnetic response of a single electromagnetic-elastic resonator by performing numerical simulations for different lengths $l_y$ of copper bricks inside the elastic media, but maintaining the total height of the unit cell, $L_S$ (cf.~Fig.~\ref{Fig-Setup}b). Consequently, the spatial configuration of the electromagnetic resonators that compose our  flat lens is obtained by imposing a hyperbolic phase profile on the metasurface,
\begin{eqnarray}
\phi(x)=\frac{2\pi}{\lambda}\left[\sqrt{x^2+f^2}-f\right]+\phi_0,
\label{Eq-PhaseProfile}
\end{eqnarray}
where $\lambda$ stands for the wavelength, which in our experiments is 3 cm, the length $f$ denotes the phase focal point which has been set to be twice the working wavelength, $f=2\lambda$, and $\phi_0$ is a homogeneous phase shift having the meaning of the phase accumulated by the wave reflected from the centre ($x=0$) of the structure. Without loss of generality, we will set $\phi_0 = 1.03$, which will fit later to the experimentally measured data.
Secondary waves emerging from different sections of the metasurface whose reflection phase is given by the Eq.~
\eqref{Eq-PhaseProfile} constructively interfere at the focal plane, similar to the waves that emerge from conventional lenses. Depending on the parameters, the phase change given by this equation can be larger than 2$\pi$ that we can provide with our metasurface, therefore we ``wrap'' the phase, so that it stays in the [0,2$\pi$] range. This introduces abrupt phase changes in certain spatial locations.
\begin{figure}[ht!]
\centerline{\includegraphics[width=\columnwidth]{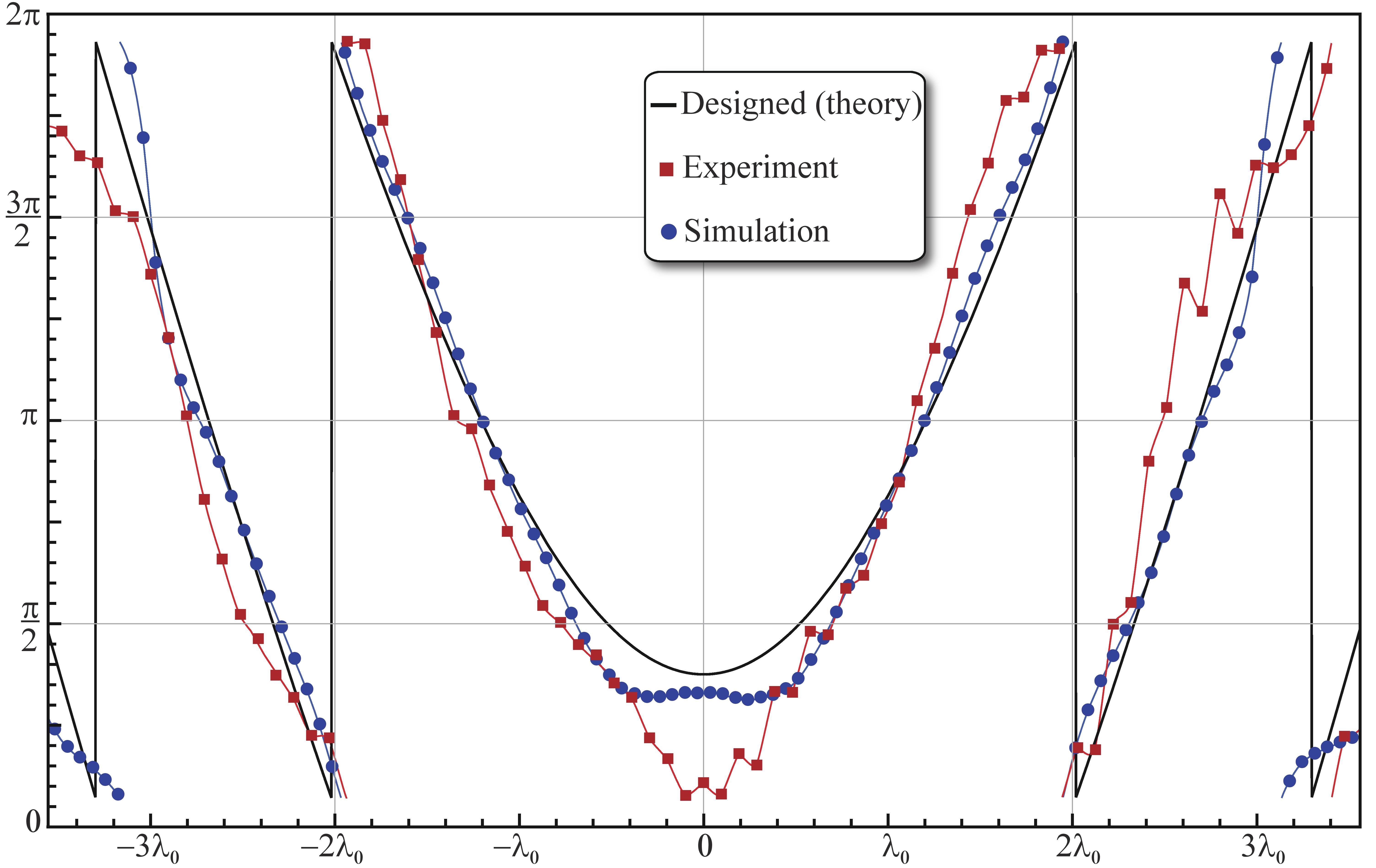}}
\caption{Comparison of phase profile of the focusing metasurface for the unstretched configuration obtained from experiment, numerical simulation and theory.}
\label{Fig-Phase}
\end{figure}
To make a metasurface that acts as a lens, the size of the resonators and their positions have been chosen to match the phase profile obtained in Eq.~\eqref{Eq-PhaseProfile}, producing a discretised version of the required phase. In Figure \ref{Fig-Phase} the ideal phase profile given by Eq.~\eqref{Eq-PhaseProfile} is contrasted with the transversal phase profile of the vertical electric field obtained from the numerical simulation and experiment measurement of the unstretched metasurface. In order to avoid the influence of the strongly inhomogeneous near field, both transversal phase profiles were obtained along a line parallel to the sample and approximately 9.5mm from the sample.

{\em Tunable Focusing.} Our simple design allows effective control over the focal length by means of strain applied to the metasurface. The underlying mechanism is straightforward; when the flexible metasurface is elongated the electric resonators are separated evenly from each other,  generating a change in the hyperbola eccentricity (cf.~ Eq.~\eqref{Eq-PhaseProfile} and Fig.~\ref{Fig-Phase}) which in turn produces a new focal distance. In order to clarify this, we rearrange Eq. \eqref{Eq-PhaseProfile},
\begin{eqnarray}
\frac{\left(\frac{\lambda}{2\pi}(\phi(x)-\phi_0)+f\right)^2}{f^2}-\frac{x^2}{f^2}=1.
\label{Eq-PhaseProfileB}
\end{eqnarray}
In terms of the hyperbola, the center is located at $p=(0,-f)$ and its  semi-major axis, $a$, as well as its  semi-minor axis, $b$, are equal to the phase focus, $a=b=f$. Therefore the eccentricity, defined as $\mathit{e}\equiv$ $\sqrt{1+\left(\frac{b}{a}\right)^2}$, and the foci of the geometrical curve (in the positive region) are given by  $\mathit{e}=\sqrt{2}$ and $F=\left(0,\left(\sqrt{2}-1\right)f\right)$, correspondingly. It is worth noting that the ideal phase profile, Eq.~\eqref{Eq-PhaseProfileB}, follows a rectangular hyperbola. Consequently, when the system is stretched, the separation between the electromagnetic resonators increases uniformly and thus also the size of the elastic metasurfaces. From the mechanical numerical simulations and empirical tests of the sample (cf.~inset in Fig.~\ref{Fig-Setup}b) it is seen that the displacement of the resonators is linearly proportional to the applied strain. Thus, after the elongation, a resonator in the position $x$ moves to $x'=\beta x$, where $\beta>1$ accounts for the strain applied to the elastic metasurface, which causes the phase profile and the focus are modified according to $\phi(x)\to\phi'(x')$ and $f\to f'$.  For the sake of simplicity, and with the aim of getting insight of how  the phase focus of the metasurface is affected by the applied strain, we have set $ \phi'(x') = \phi(x) $. In other words, we assume that the electromagnetic response of the resonator at position $x$ is not affected by the elongation of the sample and there only occurs a change in its position relative to the surface. Hereby, after stretching the eccentricity and the foci of the hyperbola read, $\mathit{e}'=\sqrt{1+\beta^2}$ and $F'=\left(0,\left(\sqrt{1+\beta^2}-1\right)f'\right)$, correspondingly. This simple calculation demonstrates that the eccentricity of the hyperbolic phase profile is modified as the elastic metasurface is stretched. Moreover, the proposed mechanism does not preserve the right angle between the asymptotes of the hyperbola, affecting the efficiency of the focusing metasurface.
\begin{figure}[ht!]
\centerline{\includegraphics[width=\columnwidth]{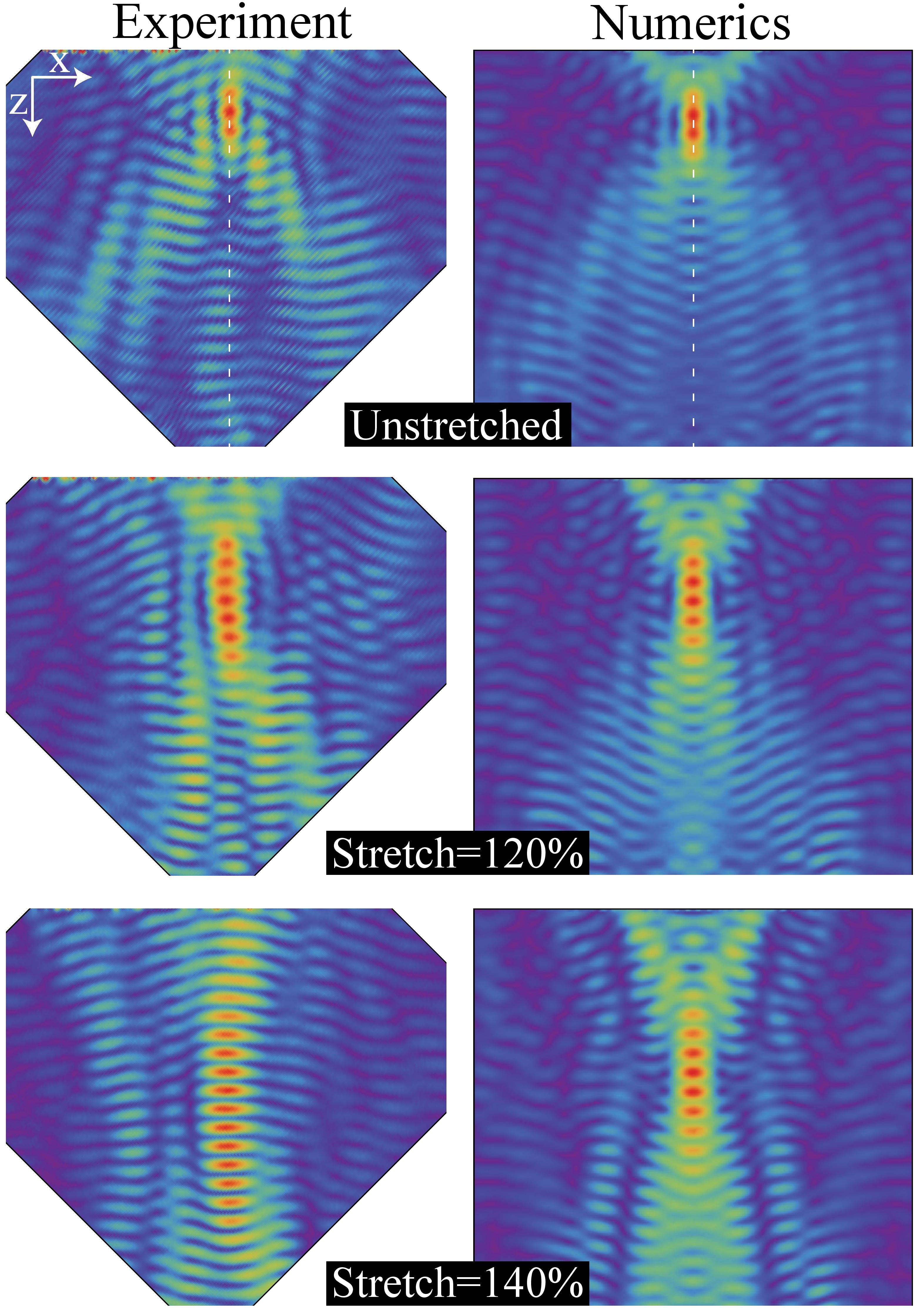}}
\caption{ Modulus of the vertical component of the scattered electric field obtained from experiment measurements (left) and numerical simulations (right) for the states unstretched (top), $120\%$ stretched (mid) and $140\%$ stretched (bottom). The metasurface is located at the top of each plot ($z=0$).  }
\label{Fig-ExpVsNum}
\end{figure}
In Figure \ref{Fig-ExpVsNum} are shown the amplitude of the vertical component of the scattered electric field obtained from experiment and numerical simulations, for different levels of strain applied to the focusing metasurface. The sample  is located at the top of each plot ($z=0$). It can be easily seen that a standing wave emerged on the system which is caused by the interaction between the ground plane behind the sample and the \emph{mechanism} that has been used to generate the wave propagating in the guide wave. In the experimental case, an antenna is placed in the focus point of a parabolic metallic mirror to produce the propagative electromagnetic wave inside the parallel plate waveguide. While in the numerical simulations a rectangular port, whose dimensions were chosen to match the cross section of the wave generated experimentally in the empty wave guide, is used. Notice that the tuning of the focusing length  is accompanied by a spherical aberration-like effect: as the sample is stretched the focal region becomes larger and  also decreases in amplitude.

{\em Spherical Aberrations}
To understand the mechanism  of tunable focusing, and thus deliver any meaning on how the phase focus changes as function of the stretching, we perform an analytical analysis.  From the stretching coefficient, $x'/x=\beta$, and Eq.~\eqref{Eq-PhaseProfileB} it is possible to obtain an expression, after introducing again the previous assumption on the phase ($\phi'(x')=\phi(x)$), that describes the dependency of the ratio of phase focus as function of the stretching,
\begin{eqnarray}
\frac{f'}{f}=\beta^2+\frac{\left(\beta^2-1\right)}{2}\left[\sqrt{1+\left(\frac{x}{f}\right)^2}-1\right].
\label{Eq-PhaseFocus}
\end{eqnarray}
Thus a linear stretching produces a nonlinear (quadratic) change in the focal length of the metasurface. Interestingly, Eq.~\eqref{Eq-PhaseFocus} reveals that the metasurface suffers of spherical aberrations in the sense that, when stretched, the electromagnetic waves reflected by the resonators located far from the center of the surface ($x\ll f$) converge in a focal distance longer than the central ones, generating an elongated focal spot. Moreover, the space-independent part of Eq.~\ref{Eq-PhaseFocus}, $f'=\beta^2 f$, gives the lower limit for the focus position as function of the stretching applied.
\begin{figure*}[ht!]
\centerline{\includegraphics[width=\textwidth]{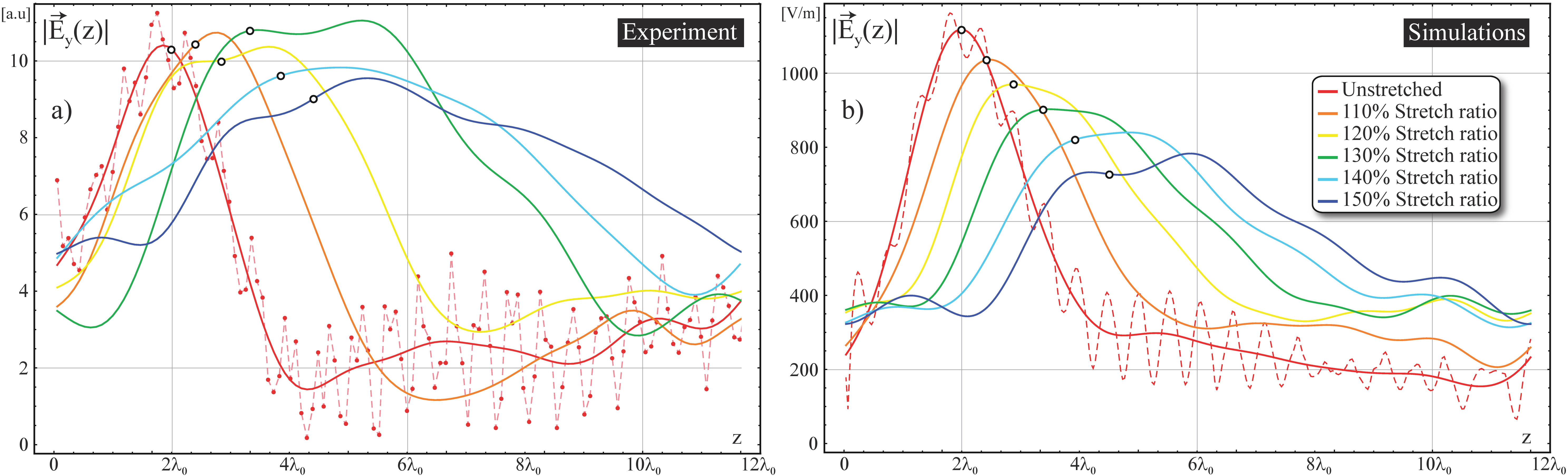}}
\caption{Filtered amplitude of the vertical component of the electric scattered field from a plane  that passes through the focus and it is perpendicular to the metasurface. Dashed lines correspond to the unfiltered signals in the unstretched state. Notice that there is an increase in the experimental amplitude (a), not observed in the numerical simulations (b), due to the interaction between the parabolic mirror used to generate the propagating electromagnetic wave and the focusing metasurface. The focus distances ($\circ$) have been obtained by taking the space-independent part of Eq.~\ref{Eq-PhaseFocus}, $f'=\beta^2 f$, which gives the lower limit for the focus position.}
\label{Fig-LineExpVsNum}
\end{figure*}
In order to determine how the focal length changes as function of the elongation applied to the proposed reflective tunable metasurface, we have extracted the field along a line perpendicular to the metasurface that passes through the focus (cf.~dashed white lines in Fig.~\ref{Fig-ExpVsNum}). Furthermore, to suppress the oscillations present in the system, and so more precisely determine the focal point, we have Fourier transformed this data. Consecutively, we applied a square window filter to eliminate the modes corresponding to the standing wave that emerged due to the experimental setup, and then we have inverse Fourier Transformed. These filtered EM fields are shown in Figure \ref{Fig-LineExpVsNum}. Notice that as the metasurface is stretched, the focus moves away from the sample. Moreover, an increase in amplitude occurs in the experiment, a phenomenon  not observed in numerical simulations, produced by the interaction of two focusing objects; our metasurface and the parabolic metallic  mirror used to generate the incident wave. Such interplay can be noticed from the experimental curve for the unstretched case, where a bump near $ 7\lambda_0 $ is observed (cf.~Fig.~\ref{Fig-LineExpVsNum}a). Thus, when the metasurface is stretched, the focal length  approaches this region producing a collective effect which increases the focused beam amplitude. On the other hand,  this coupling effect is not observed in the numerical simulations because of a rectangular port used to emulate excitation. It is therefore expected that if another method is used to generate the electromagnetic wave inside the parallel plate wave guide, a decrease in the amplitude occurs as shown in the numerical simulations (cf.~Fig.~\ref{Fig-LineExpVsNum}b). We obtained the positions of the focus (cf.~ black circles in Fig.~\ref{Fig-LineExpVsNum}) by only considering the space-independent part of Eq.~\ref{Eq-PhaseFocus}, $f'=\beta f$, which represents the lower limit for the focus position.

In conclusion, a tunable elastic metasurface lens able to change its focal length for linear polarized electromagnetic wave is proposed and experimentally demonstrated. The measurements are in agreement with the theory and numerical simulations. Our design provides a simple and effective way to  externally tune the functionality of the metasurface by means of stretching. The adjustable focus mechanism is based on the change of eccentricity suffered by the hyperbolic profile phase of the metasurface while being tensioned. Thereby, a simple analysis of the shape of the hyperbola reveals that in the ideal case (unstretched metasurface), the required phase profile must be a rectangular hyperbola but in the proposed tuning mechanism this is not preserved after stretching. Therefore, in order to maintain the maximum efficiency of the metasurface it is necessary to exploit other degrees of freedoms so that, regardless of the stretching applied and the new focusing distance after the elongation, the corresponding phase profile satisfies Eq.~\eqref{Eq-PhaseProfile}. Although there are some discrepancies between the experiment and the numerical results, they have been discussed and understood in their entirety. Our proposal demonstrates that it is possible to use elastic smart metamaterials as a platform for new tunable devices for electromagnetic beam steering and shaping.


{\em Acknowledgments} D.A.P. acknowledge support from the Australian Research Council under the Discovery program DP150103611. Y. Z. acknowledges the financial support
of CONICYT by BecasChile 72130436/2013.

\end{document}